\documentclass[letter]{aa}
\usepackage{graphicx}
\usepackage[authoryear]{natbib}
\usepackage{amsmath}
\usepackage{amssymb}
\usepackage{txfonts}
\usepackage{threeparttable}
\begin{document}
   \title{Quasi-periodic flares in EXO\,2030+375 observed with INTEGRAL}

   \author{D. Klochkov
          \inst{1}
          \and
          C. Ferrigno\inst{2}
          \and
          A. Santangelo\inst{1}
          \and
          R. Staubert\inst{1}
          \and
          P. Kretschmar\inst{3}
          \and
          I. Caballero\inst{4}
          \and
          K. Postnov\inst{5}
          \and
          C.~A.~Wilson-Hodge\inst{6}
          }

   \institute{Institut f\"ur Astronomie und Astrophysik, Universit\"at
     T\"ubingen (IAAT), Sand 1, 72076 T\"ubingen, Germany
     \and
     ISDC Data Center for Astrophysics of the University of Geneva
     chemin d'\'Ecogia, 16 1290 Versoix, Switzerland
     \and
     European Space Astronomy Centre (ESA/ESAC), Science Operations
     Department, Villanueva de la Can\~ada (Madrid), Spain 
     \and 
     CEA Saclay, DSM/IRFU/SAp-UMR AIM (7158) CNRS/CEA/Universit\'e P. Diderot--F-91191
     \and
     Sternberg Astronomical Institute, Universitetski pr. 13, Moscow, 119991, Russia
     \and
     NASA Marshall Space Flight Center, Huntsville, AL 35812, USA
   }

   \date{Received ***; accepted ***}
   
   \abstract
   % context heading (optional)
   {
     Episodic flaring activity is a common feature of X-ray pulsars
     in HMXBs. In some Be/X-ray binaries flares were observed in
     quiescence or prior to outbursts. EXO\,2030+375 is a Be/X-ray binary 
     showing ``normal'' outbursts almost every $\sim$46 days,
     near periastron passage of the orbital revolution. 
     Some of these outbursts were occasionally monitored with 
     the INTEGRAL observatory.
   }
   % aims heading (mandatory)
   {
     The INTEGRAL data revealed strong quasi-periodic flaring activity 
     during the rising part of one of the system's outburst. Such activity
     has previously been observed in EXO\,2030+375 only once, in 1985
     with EXOSAT. (Some indications of single flares have also been observed
     with other satellites.)
   }
   % methods heading (mandatory)
   {
     We present the analysis of the flaring behavior of the source 
     based on INTEGRAL data and compare it with the flares observed 
     in EXO\,2030+375 in 1985. 
   }
   % results heading (mandatory)
   {
     Based on the observational properties of the flares, we argue that the 
     instability at the inner edge of the accretion disk is the most 
     probable cause of the flaring activity.
   }
   {}
   

   \maketitle
%
%________________________________________________________________

\section{Introduction}

Accreting X-ray pulsars, mostly residing in high-mass X-ray binaries (HMXBs),
often show abrupt increases in their X-ray luminosity lasting from
a few tens of seconds to several hours -- X-ray flares. Flaring activity
is often observed on top of a slower flux variation related to
X-ray outbursts or super-orbital modulation. Among the best-known 
``flaring'' X-ray pulsars are LMC\,X-4 \citep[e.g.][end references
therein]{Moon:etal:03}, SMC\,X-1 \citep{Angelini:etal:91,Moon:etal:03b},
Vela\,X-1 \citep{Kreykenbohm:etal:08,Fuerst:etal:10}. X-ray binaries with
a donor star of Be (or Oe) spectral type (Be/X-ray binaries or BeXRB) 
are currently the most numerous class of HMXBs with X-ray 
pulsars \citep{Liu:etal:06}, although another class of HMXB pulsars,
the supergiant fast X-ray transients (SFXT), is rapidly growing 
\citep[e.g.][]{Sidoli:11}. BeXRBs are characterized by periodic or sporadic
X-ray outbursts lasting from several days to several weeks when the
neutron star accretes matter from the equatorial disk around the 
donor star \citep[see e.g.][for a recent review]{Reig:11}.
These systems are also known
to show occasional X-ray flares \citep{Finger:etal:99,Reig:etal:08,Caballero:etal:08,
Postnov:etal:08}.

Probably the most remarkable flaring activity among BeXRBs was exhibited
by the 42\,s pulsar \object{EXO\,2030+375}. %\citep{Parmar:etal:89a}. 
A series of six roughly equidistant flares with a mean
period of $\sim$four hours was observed with the EXOSAT satellite 
a few months after the giant outburst of the source in 1985
\citep{Parmar:etal:89a}.
EXO\,2030+375 is one of the most regularly monitored BeXRBs. 
In addition to
two ``major'' outbursts in 1985 and 2006 with a peak X-ray luminosity
of $L_{\rm X}\gtrsim 10^{38}$\,erg\,s$^{-1}$, assuming a distance of 7\,kpc 
\citep{Wilson:etal:02}, the source exhibited less
powerful ``normal'' outbursts with 
$L_{\rm X}\sim 10^{37}$\,erg\,s$^{-1}$ almost every orbit. The orbital
period of the system is $\sim$46\,d \citep{Wilson:etal:08}.

Some flaring of EXO\,2030+375 has been reported since the
EXOSAT observations, but it has generally appeared to be a single 
flare during the rise of normal outbursts 
\citep{Reig:Coe:98,Camero:etal:05}. Flares were also apparently present 
during and between the normal outbursts shortly after the 2006 major outburst 
\citep[Fig. 3 in][]{Wilson:etal:08}. But the short duration and 
broad spacing of the observations make it unclear whether this is the same 
quasi-periodic phenomenon that was seen with EXOSAT.
In this Letter we present the INTEGRAL observations of the source
that, for the first time since the EXOSAT observations, reveal 
strong quasi-periodic flaring behavior.

\section{Observations and data processing}

Since its major outburst in 2006, EXO\,2030+375 has repeatedly appeared in the
field of view of \textit{International Gamma-Ray Astrophysics Laboratory}
(INTEGRAL, \citealt{Winkler:etal:03}) during the
observational program concentrated on the ``Cygnus Region''.
The INTEGRAL observatory has three main X-ray 
instruments: (i) the imager IBIS sensitive from $\sim$20\,keV to a
few MeV with high spatial and moderate spectral resolution
\citep{Ubertini:etal:03}; (ii) the spectrometer SPI, which is
sensitive in the roughly 
same energy range as IBIS, but with much higher spectral resolution and
substantially lower imaging capability \citep{Vedrenne:etal:03};
and (iii) the X-ray monitor JEM-X with moderate spectral and spatial 
resolution, operating between $\sim$3 and $\sim$35\,keV \citep{Lund:etal:03}.
INTEGRAL observations normally consist of a series of pointings called
Science Windows, 2 to 4\,ksec each.

%%%%%%%% FIGURE %%%%%%%%%%%%%%%%%
\begin{figure}
\centering
\resizebox{\hsize}{!}{\includegraphics{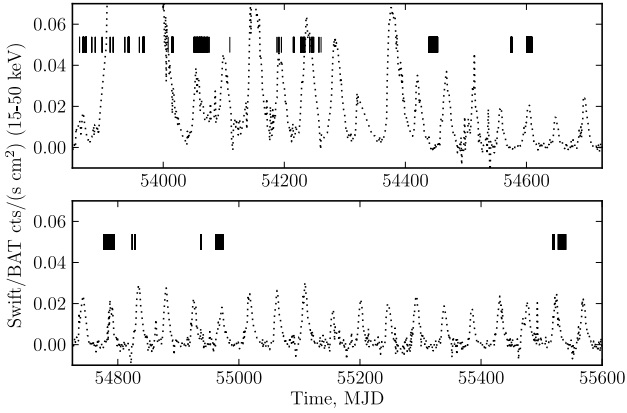}}
\caption{INTEGRAL observations of EXO\,2030+375 (vertical bars) 
%on top of 
superposed on the Swift/BAT
light curve of the source since its major outburst in 
June--October 2006.
}
\label{fig:lcobs}
\end{figure}
%%%%%%%%%%%%%%%%%%%%%%%%%%%%%%%%%

Because the times of the observations were not optimized for EXO\,2030+375, 
the INTEGRAL coverage of the source's normal outbursts is rather sparse. 
The Swift/BAT light curve of the pulsar\footnote{We used the Swift/BAT 
transient monitor results provided by the Swift/BAT team}
with the indicated INTEGRAL observations is shown in Fig.\,\ref{fig:lcobs}.

For our analysis we used the ISGRI detector layer of IBIS sensitive in the
20--300\,keV energy range \citep{Lebrun:etal:03} and JEM-X. 
Owing to limited count-rate statistics, no additional information
could be gained from SPI data.
The standard data processing was performed with 
version 9 of the Offline Science Analysis (OSA) software
provided by the INTEGRAL Science Data Centre 
(ISDC, \citealt{Courvoisier:etal:03}). We performed an 
additional gain correction of the ISGRI energy scale based
on the background Tungsten spectral lines.

\begin{figure}
\resizebox{0.9\hsize}{!}{\includegraphics{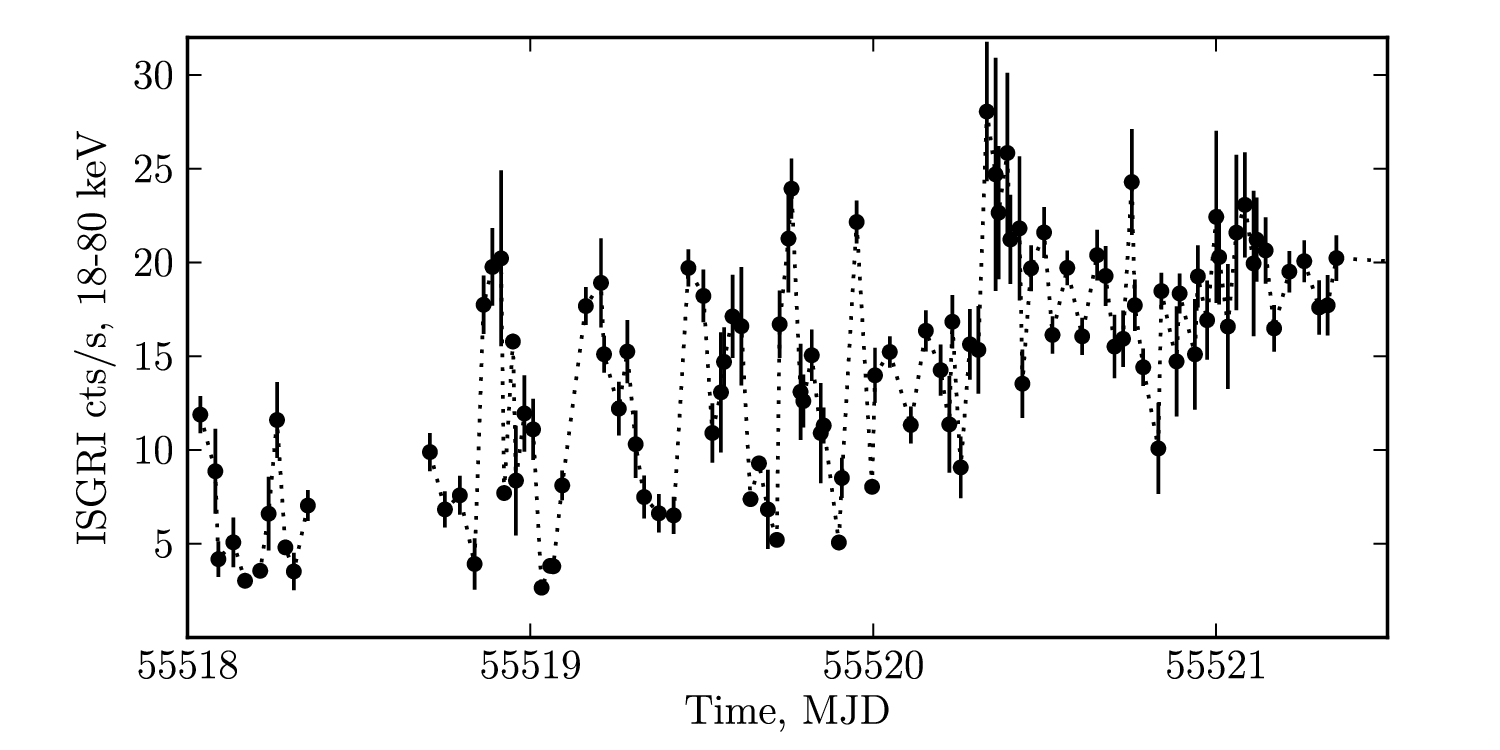}}
\caption{The INTEGRAL/ISGRI light curve of EXO\,2030+375 during the
rise of its normal outburst in November--December 2010.
The flaring activity is clearly seen. 
The light curve of the entire
outburst is shown in Fig.\,\ref{fig:nv}.
One Crab corresponds to $\sim$260 cts/s in the specified energy range.}
\label{fig:lcflares}
\end{figure}

To search for the flaring activity, we examined the entire IBIS/ISGRI 
light curve by combining the publicly available ISGRI data products
in the HEAVENS data base\footnote{http://www.isdc.unige.ch/heavens/heavens}
with the results of our own analysis. We did not find any clear
evidence of flares in all the data except for the latest
INTEGRAL observations of the source
in November and December 2010 (MJD $\sim$55520--55540),
partially covering a normal outburst.
The rising part of the outburst is shown in Fig.\,\ref{fig:lcflares}.
The upper panel of Fig.\,\ref{fig:nv} shows the entire 
outburst as observed with INTEGRAL.
In this work we concentrate on the analysis of the source's flaring 
behavior during this outburst.

%%%%%%%% FIGURE %%%%%%%%%%%%%%%%%
%%%%%%%%%%%%%%%%%%%%%%%%%%%%%%%%%

\section{Timing analysis}

As can be seen in Fig.\,\ref{fig:lcflares}, the flux in 
the rising part of the outburst experiences quasi-periodic 
oscillations/flares that cease as the averaged flux increases.
One can identify at least five subsequent flares with a 
mean period of $\sim$0.3 days ($\sim$7 hours). 

%%%%%%%%% FIGURE %%%%%%%%%%%%%%%%%
\begin{figure}
\resizebox{0.95\hsize}{!}{\includegraphics{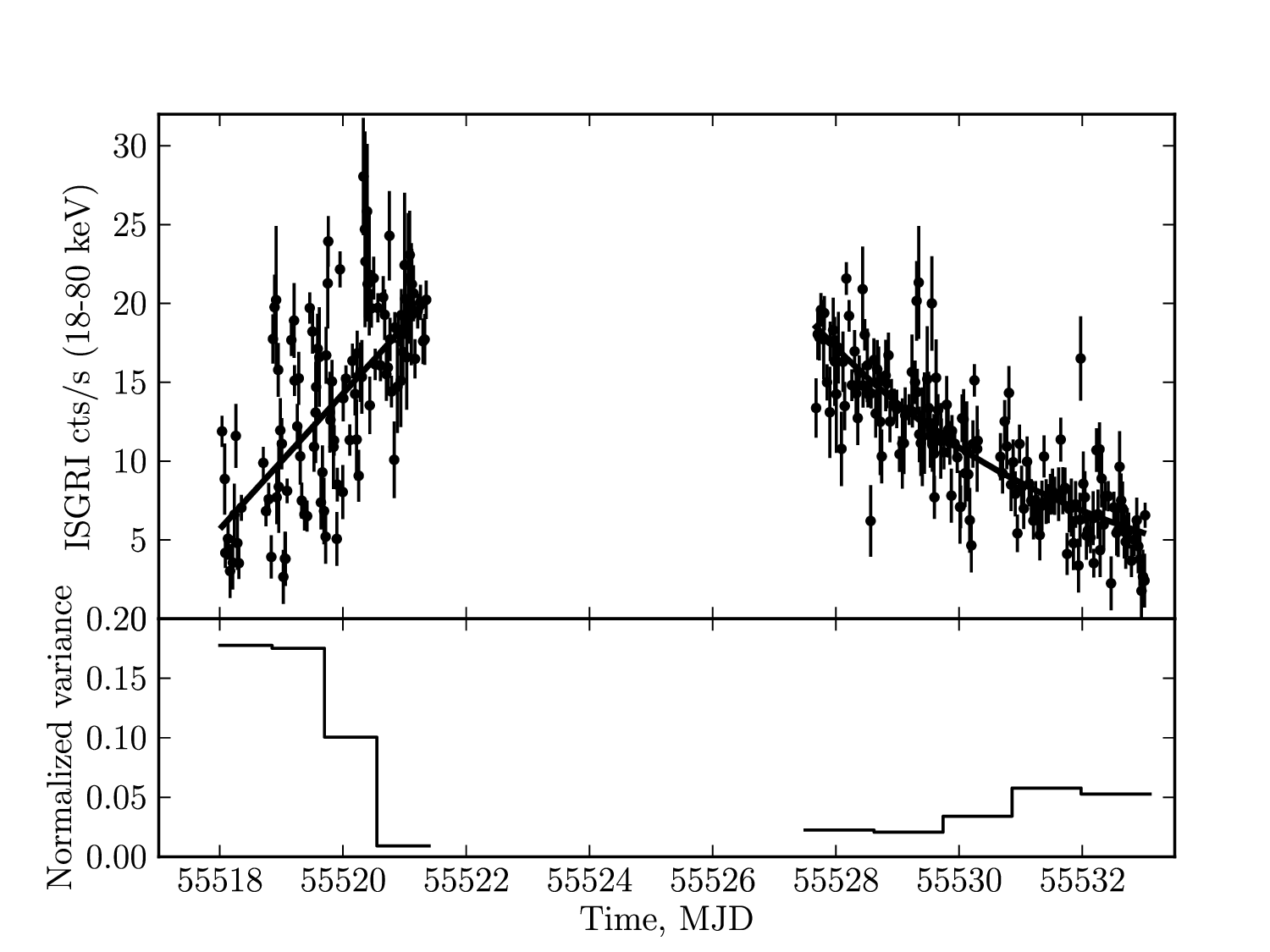}}
\caption{\emph{Top}: ISGRI light curve of the EXO\,2030+375 
outburst approximated 
with a polynomial function (solid curve) which represents 
the averaged evolution of the flux. 
\emph{Bottom:} Normalized excess variance of the source flux with respect to the
polynomial function. 
%The normalized variance represents the level of 
%intrinsic variability of the source.
}
\label{fig:nv}
\end{figure}
%%%%%%%%%%%%%%%%%%%%%%%%%%%%%%%%%

To characterize the level of the flux variability during the outburst we 
calculated the normalized excess variance in the light curve in relatively
broad time intervals.
We defined four equal adjacent intervals in the rising part and five
equal adjacent intervals in the decay.
The normalized excess variance is often used as a simple measure of the 
intrinsic variability amplitude in light curves, 
\citep[see e.g.][]{Nandra:etal:97}:
\begin{equation}
  \sigma^2_{\rm NXS} = \frac{1}{N\langle f\rangle^2}\Sigma_{i=1}^{N}[(f_i-f_i^{\rm aver})^2 - \sigma_i^2].
\end{equation}
Here $N$ is the number of data points in the 
corresponding time interval, 
%$f_i$ and $\sigma_i$ -- flux
%and uncertainty of the individual data points respectively, 
$f_i$ is the flux of the individual data points, $\sigma_i$ -- their
uncertainty, $f_i^{\rm aver}$
is the smoothed evolution of the flux obtained by a polynomial fit to 
the light curve (upper panel of Fig.\,\ref{fig:nv}), and
$\langle f\rangle$ is the mean value of the flux within the interval.
The normalized variance $\sigma^2_{\rm NXS}$ 
in our case represents the amplitude of
intrinsic flux variations superimposed on the smoothed flux
development. The term $\sigma_i$ under the summation ($\Sigma$) eliminates
the contribution of the Poisson noise.
The bottom panel of Fig.\,\ref{fig:nv} represents the normalized
variance as a function of time during the outburst of EXO\,2030+375.
It can be seen that the amplitude of the variability is high
at the rising phase (corresponding to the flaring episode). Then it decreases
towards the maximum of the outburst and remains low during the decay.

\begin{figure}
\resizebox{0.95\hsize}{!}{\includegraphics{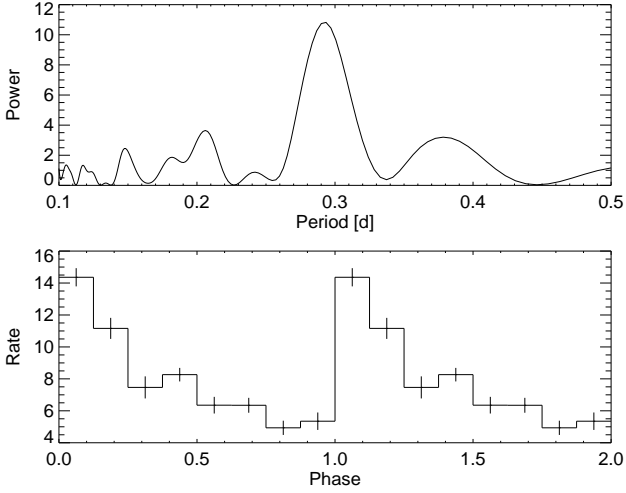}}
\caption{\emph{Top}: The Lomb-Scargle periodogram of the ``flaring'' part of 
EXO\,2030+375 light curve (between MJD 55518.5 and 55521.5). 
The peak around $\sim$0.3\,d is clearly seen.
\emph{Bottom:} The ``flaring'' part of the light curve folded with best period
found from the periodogram (0.293\,d).
}
\label{fig:period}
\end{figure}

To study periodicity of the flares, we performed a formal period search
in the rising part of the outburst (between MJD 55518.5
and 55521.5) using the Lomb-Scargle periodogram. 
The results are presented in Fig.\,\ref{fig:period}. The periodogram shown
in the top panel indicates a clear peak around $\sim$0.3\,days ($\sim$7\,hours).
The averaged profile obtained by folding the light curve with this period
is shown in the bottom panel. The profile shape is 
asymmetric and characterized by a steep rise and a slower decay. 

%%%%%%%%% FIGURE %%%%%%%%%%%%%%%%%
%%%%%%%%%%%%%%%%%%%%%%%%%%%%%%%%%

\begin{figure}
\resizebox{0.9\hsize}{!}{\includegraphics{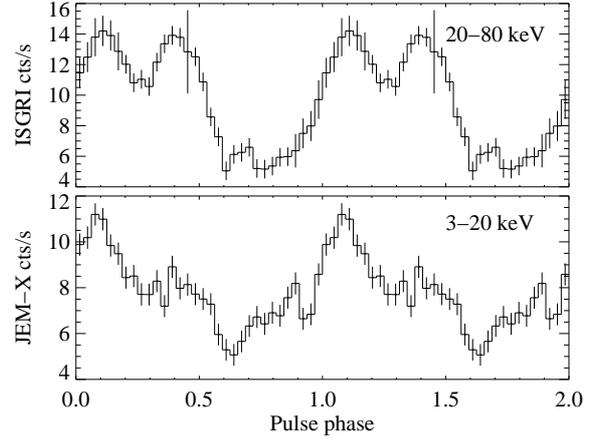}}
\caption{Pulse profiles of EXO\,2030+375 obtained with ISGRI (top) and
  JEM-X (bottom) instruments during the outburst (MJD 55518.5 -- 55533.0).
}
\label{fig:pp}
\end{figure}

We used the INTEGRAL data to study the pulse period behavior and pulse
profiles during the outburst. The photon arrival times were converted
to the reference frame 
of the solar system barycenter and corrected for the binary orbital
motion using the ephemeris by \citet{Wilson:etal:08}. Using the
pulse-phase-connection technique 
\citep[e.g.][]{Staubert:etal:09}, 
we found the pulse period $P=41.31516(2)$\,s at the
epoch $T_0$(MJD)$=55526.056994$ and the period derivative $\dot P
=-1.9(1)\times 10^{-9}$\,s/s, that indicates significant spin-up. We used
the measured pulse ephemeris to construct and study the pulse
profiles of the source. We could not find any
difference between the profiles obtained during the rise and decay of
the outburst. Figure\,\ref{fig:pp} shows the profiles accumulated
during the entire outburst.

%%%%%%%%% FIGURE %%%%%%%%%%%%%%%%%
%%%%%%%%%%%%%%%%%%%%%%%%%%%%%%%%%

\section{Spectral analysis\label{sec:spe}}

For the spectral analysis we used JEM-X data between 3.5 and 35\,keV,
and ISGRI data between 20 and 80\,keV. We added systematic uncertainties
at the level of 2\% to the \textit{JEM-X} spectra and 1\% to the
\textit{ISGRI} spectra based on the recommendations of the
instrument teams and the Crab observations. 
The spectrum of the source during the outburst 
(accumulated between MJD 55518.5 and 55533.0) was modeled with the 
cutoff-powerlaw model ($F(E)\propto E^{-\Gamma}\times\exp{[E/E_{\rm fold}]}$,
where $E$ is the photon energy, $\Gamma$ and $E_{\rm fold}$ are the photon
index and the folding energy, respectively) modified at lower energies
by photoelectric absorption. 

The best-fit parameters are
$\Gamma=1.6(1)$, $E_{\rm fold}=30(2)$\, keV, the absorption column density
$n_{\rm H}=11(1)\times 10^{22}$
hydrogen atoms per cm$^2$. The uncertainties in
parentheses refer to the last digit and are at the 90\% confidence level.
The value of $n_{\rm H}$ is substantially higher than
measured in previous observations of the source, including
older INTEGRAL measurements 
\citep[1--3$\times 10^{22}$\,cm$^{-2}$, e.g.][]{Klochkov:etal:07, 
Wilson:etal:08}.
We note, however, that studying $n_{\rm H}$ with JEM-X is generally 
problematic as the data below $\sim$3\,keV are not available.
The significance of the measured increase in absorption is 
therefore questionable.

The 
%photon statistics 
data quality does not permit spectroscopy
of individual flares. To characterize the spectral
behavior of EXO\,2030+375 during the flaring episode and compare
it with the rest of the outburst, we explored the luminosity-dependence
of the source spectrum during the rising (flaring) part and the decay of
the outburst. We grouped the individual INTEGRAL pointings 
according to the measured flux
in the 20--80\,keV range. For each group we extracted and analyzed
the X-ray spectrum using the spectral model described above.  
The statistics did not allow us to explore the dependence of each
individual spectral parameter on flux. We therefore fixed $n_{\rm H}$ and
$E_{\rm fold}$ to their averaged values ($11\times 10^{22}$\,cm$^{-2}$
and 30\,keV, respectively) and explored the photon index $\Gamma$
as a function of flux. 

%%%%%%%%% FIGURE %%%%%%%%%%%%%%%%%
\begin{figure}
\resizebox{0.95\hsize}{!}{\includegraphics{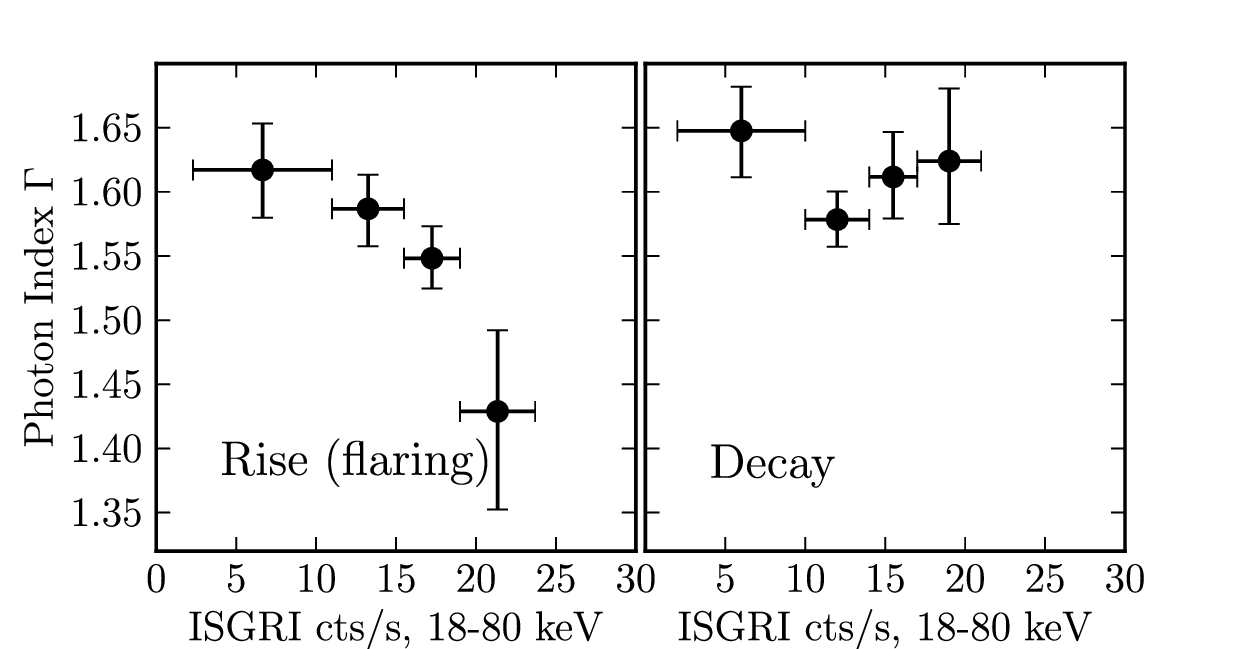}}
\caption{Photon index $\Gamma$ as a function of flux during the
rising (flaring) part of the outburst \emph{(left)} and the decay
phase \emph{(right)}.
One Crab corresponds to $\sim$260 cts/s in the specified energy range.} 
\label{fig:frs}
\end{figure}
%%%%%%%%%%%%%%%%%%%%%%%%%%%%%%%%%

The resulting dependence is shown in Fig.\,\ref{fig:frs} for the
rise (left) and decay (right) of the outburst. Higher values of $\Gamma$
correspond to a softer spectrum (see the model description above). 
The horizontal error bars represent the width of the flux bins. 
The vertical error bars indicate $1\sigma$-uncertainties.
While the spectrum apparently tends to get harder with 
increasing flux during the rising part of the outburst, 
it remains roughly constant during the decay.
To quantify this behavior, we performed linear fits to the data points
and calculated the slopes with the corresponding uncertainties. 
The slope is 
$(-9.0\pm3.7)\times 10^{-3}$\,(cts/s)$^{-1}$ in the rising part and
$(2.8\pm 4.2)\times 10^{-3}$\,(cts/s)$^{-1}$ in the decay of the outburst 
(uncertainties at 1$\sigma$ confidence level).

\section{Discussion}

A direct comparison of the flares presented here with those observed
with EXOSAT in 1985 revealed significant similarity. 
The peak fluxes and the relative amplitude of the flares are roughly
the same in both episodes. Also the average shape -- fast rise / slow
decay -- is similar in the two cases. The mean period is, however, 
different: $\sim$4\,hr for the EXOSAT flares and $\sim$7\,hr for the
INTEGRAL ones. The location of the 
EXOSAT flares with respect to the nearest ``normal'' outburst 
is difficult to reconstruct because no
monitoring of the source flux (apart from the EXOSAT data themselves) was
performed at the time. 
Since the orbital phase of ``normal'' 
outbursts varies significantly with time \citep{Wilson:etal:02},
the extrapolation of the orbital phase ephemeris back to the EXOSAT 
observations would not resolve the problem.

X-ray flares in accreting pulsars are usually attributed to one of the 
following mechanisms: (1) instabilities of the accretion flow around/within
the magnetospheric boundary \citep[e.g.][]{Moon:etal:03,Postnov:etal:08}, 
(2) highly inhomogeneous stellar wind  of the donor star 
\citep[e.g.][]{Taam:etal:88,
Walter:etal:07}, and (3) nuclear burning at 
the neutron star surface \citep[e.g.][]{Levine:etal:00,Brown:Bildsten:98}.
The nuclear burning scenario is, however, difficult to reconcile with
a relatively high accretion rate before and after a flare, which would suppress
the thermonuclear instability \citep{Bildsten:Brown:97}. 
In the case of EXO\,2030+375, inhomogeneities of the companion's stellar
wind are also unlikely to be a direct cause of the flares for the 
following reasons. First, the viscous time of the accretion disk
that is believed to be present in EXO\,2030+375 during normal outbursts
\citep{Wilson:etal:02} and to even survive during quiescence
\citep{Hayasaki:Okazaki:06}, is at least several days, which would
smooth out any variations in the mass accretion rate $\dot M$ caused by
inhomogeneity of the wind shorter than this time. Second, nonuniform
stellar wind cannot explain the observed quasi-periodic appearance 
of the flares. On the other hand, various kinds of
magneto-hydrodynamic instabilities 
at the inner edge of the accretion disk may easily lead to oscillations
in the mass flow towards the polar caps of the neutron star
\citep{Apparao:91,Postnov:etal:08,DAngelo:Spruit:10}, leading to the
observed flaring activity. For example, \citet{DAngelo:Spruit:10} have
illustrated that when the magnetospheric radius $r_m$ (where magnetic
field of the neutron star truncates the accretion disk) is larger but close
to the corotation radius $r_c$ (where the Keplerian frequency is
equal to the accretor's spin frequency), matter in the inner regions
of the disk will pile up leading to an increase in the local gas
pressure and, therefore, a decrease in $r_m$. When $r_m$ crosses
$r_c$, the accumulated reservoir of gas is accreted by the neutron
star and the cycle repeats.

To assess the applicability of this scenario to the flares observed in
EXO\,2030+375, we estimated the expected characteristic time scale 
of $\dot M$-oscillations. Without going into physical details of the
disk-field coupling at $r_m$ one would generally expect that the
oscillations in $\dot M$ take place on the time scale close to
the local viscous time at $r_m\sim r_c$. Following the standard
$\alpha$-viscosity prescription \citep{Shakura:Sunyaev:73}, this time
can be estimated as $\tau_c\sim r_c^2/\nu(r_c)\sim 
1/[\Omega\alpha(H/R)^2]$, where $\nu(r_c)$ 
is the viscosity at $r_c$, $\Omega$ is the spin frequency of the
neutron star, and $(H/R)$ the semithickness of the accretion
disk. Using ``standard'' values of $\alpha=0.1$, $H/R=0.05$, and the
known pulse period $P\simeq 40$\,s, one gets $\tau_c\sim 7$\,hr,
i.e. of several hours, as was observed. 

The averaged profile of the flares is characterized by a steep rise and
a slower decay (Fig.\,\ref{fig:period}), which is very similar
to the flares observed with EXOSAT in 1985 (Fig.\,2 of 
\citealt{Parmar:etal:89a}). According to the authors, such a shape 
suggests a ``draining reservoir'' that is in line with the picture
described above (matter piling up on the inner edge of the disk).

The observed difference in the spectral behavior between the 
flaring part and the rest of the outburst (Sect.\,\ref{sec:spe})
suggests different configurations of the region where matter
couples to the field lines. Such a difference is indeed expected
if the flares are caused by the oscillating inner edge of the accretion
disk. In this case, matter from the oscillating inner disk rim 
would couple to different dipole field lines
of the neutron star (and follow them) compared to the decay part of 
the outburst where the configuration of the inner disk rim
is presumably stable.

The difference in the mean period of flares in the INTEGRAL
and EXOSAT observations can also be understood in the described picture.
The period must depend on the mass transfer rate through the accretion
disk, i.e. time needed to refill the reservoir. This rate could be
different between the EXOSAT and INTEGRAL observations due to,
e. g., changes in the state of the Be-disk. 
In Fig.\,\ref{fig:lcflares}
one might also notice some shortening of the the flare separation
time as the flux increases (although this behavior is difficult to
quantify with the available statistics). Such behavior, if real,
might reflect shortening of the reservoir refill time as the mass
transfer rate increases towards the maximum of the outburst.

Thus, we argue that the observational appearance of the flares
in EXO\,2030+375 suggests that the instability of the inner disk edge 
(pile-up/draining of matter) is the most probable cause of the flares.

It is important to note that 
the rarity of the detected flaring episodes 
(even considering the relatively sparse observational coverage) 
means that the range of physical conditions needed to initiate
flares could be very narrow, which would 
lead to the serendipitous character of the phenomenon. 

\begin{acknowledgements}
The work was supported by the Carl-Zeiss-Stiftung and by DLR grant BA5027.
This research is based on observations with INTEGRAL, an ESA 
project with instruments and science data centre funded by ESA member 
states. 
The authors thank the anonymous referee for useful suggestions.
\end{acknowledgements}

\bibliographystyle{aa}
\bibliography{ref}

\end{document}